\documentclass[trackchanges]{aastex701}

\usepackage{amsmath}

\begin{document}

\title{Age and Mass Signatures in the Multiband Second-order Image Structure of Young Star Clusters in M31}

\author[orcid=0009-0003-5731-3761,gname='Yuan', sname='Liang']{Yuan Liang}
\affiliation{School of Astronomy and Space Science, University of Chinese Academy of Sciences, Beijing 100049, People's Republic of China}
\email[show]{liangyuan@bao.ac.cn}  

\author[orcid=0000-0002-9390-9672, gname='Chao-Wei', sname='Tsai']{Chao-Wei Tsai} 
\affiliation{School of Astronomy and Space Science, University of Chinese Academy of Sciences, Beijing 100049, People's Republic of China}
\affiliation{National Astronomical Observatories, Chinese Academy of Sciences, Datun Road A20, Beijing, People's Republic of China}
\affiliation{Institute for Frontiers in Astronomy and Astrophysics, Beijing Normal University, Beijing 102206, People's Republic of China}
\email{cwtsai@nao.cas.cn}

\author[orcid=0000-0001-7808-3756, gname=Jingwen,sname=Wu]{Jingwen Wu}
\affiliation{School of Astronomy and Space Science, University of Chinese Academy of Sciences, Beijing 100049, People's Republic of China}
\affiliation{National Astronomical Observatories, Chinese Academy of Sciences, Datun Road A20, Beijing, People's Republic of China}
\email{jingwen@nao.cas.cn}

\begin{abstract}

The ages and masses of star clusters may leave systematic signatures in the second-order spatial structure of semi-resolved photometric images. To test this hypothesis, we analyze the images using the Hessian matrix and identify basic luminous structures, termed peak cores. Applying this analysis to six-band PHAT/AP images of 1249 M31 clusters with CMD-based ages of 10--300 Myr and masses of $10^{2.2}$--$10^{4.5}\,M_{\odot}$, we examine the correlations of age and mass with the peak-core structural parameters, $C_S$ and $Q$. Younger clusters show larger $C_S$ values in the ultraviolet bands, with the strongest age--$C_S$ anti-correlation in F336W ($\rho=-0.61$), whereas more massive clusters show smaller $Q$ values, most clearly in F336W ($\rho=-0.44$). These correlations suggest that the contrast of UV-bright peak cores decreases with age, whereas increasing mass alters the abundance and spatial organization of detectable structures across different bands.

\end{abstract}




\section{Introduction} \label{sec:intro}

Star clusters are fundamental laboratories for studying star formation and the early dynamical evolution of stellar systems. Yet many questions about their early evolution remain unresolved, including how strongly newly formed clusters inherit the hierarchical structure of their natal environments, how long such primordial structure survives, and how young systems evolve toward smoother and more centrally organized configurations \citep{longmore2014formation, krumholz2019star, renaud2018star, adamo2020star}. Researchers have often used cluster structure to constrain these questions. For example, \citet{arnold2024kinematic} investigated the inheritance of natal structure through internal kinematic substructure; \citet{sanchez2009spatial} examined the persistence of spatial substructure with age; \citet{dellacroce2024} traced early dynamical evolution through cluster expansion; and \citet{tarricq2022structural} related longer-term evolution to changes in core and halo structure.

Nevertheless, most structural studies of star clusters have inferred their physical properties in one of two limiting observational regimes: resolved systems, where the spatial and kinematic organization of individual stars can be characterized, and unresolved systems, where clusters are described through their integrated light or relatively simple photometric profiles. Between these two regimes lies the semi-resolved case, in which individual cluster members cannot be fully recovered but substantial internal structure remains visible.

How cluster physical properties are manifested in the semi-resolved regime has received comparatively limited attention. The few existing studies have shown that pixel-scale photometric variations can constrain the age information of semi-resolved stellar populations \citep{cook2019measuring, cook2020measuring}, including the ages of star clusters \citep{whitmore2011using}, while internal morphology has been used for cluster classification \citep{whitmore2021star, thilker2022phangs, deger2022bright}. More recently, \citet{miller2026objective} characterized cluster structure through hierarchically nested isophotes. In general, these studies remain relatively fragmented, and the quantities they employ are still primarily derived from the surface-brightness distribution itself.

In recent years, high-resolution surveys of nearby galaxies have greatly expanded the number of star clusters observed in the semi-resolved regime. Thousands of clusters in M31 and M33 provide particularly favorable samples for probing semi-resolved internal structure \citep{johnson2015phat, johnson2022phatter}, while the $\sim10^5$ clusters and compact associations identified across 38 nearby galaxies by PHANGS-HST, although observed at somewhat lower physical resolution, constitute a valuable resource for future statistical studies \citep{maschmann2024phangs}. More importantly, the galaxy-wide coverage of these samples makes it possible not only to investigate how cluster physical properties are reflected in semi-resolved structure, but also to establish connections between cluster-scale structure and the larger-scale galactic environment. These opportunities make it increasingly important to develop a systematic framework specifically designed to describe and quantify star-cluster structure in the semi-resolved regime.

We describe a semi-resolved cluster as a continuous photometric field and propose that its physical properties may be encoded not only in the intensity distribution, i.e., the zeroth-order information, but also in its second-order spatial structure. Our recent work \citep{liang2026hessian} supports this idea by showing a significant correlation between cluster age and a global measure of curvature variation derived from the Hessian matrix. This result motivates us to explore the local structural units that more fundamentally characterize the second-order organization of the cluster light field.

In this work, we identify these local structures as \emph{peak cores}, defined as connected regions where the Hessian is negative definite and representing luminous entities in the cluster light field. Based on the curvature and spatial organization of these units, we construct two parameters, $C_S$ and $Q$, to quantify curvature diversity and spatial organization, respectively. Applying these diagnostics to young clusters in M31, we find that $C_S$ is more strongly associated with age, whereas $Q$ is more closely related to mass, suggesting that age and mass are preferentially linked to distinct aspects of the second-order photometric structure.

The paper is organized as follows. In Section~\ref{sec:data_method}, we describe the cluster sample and introduce the peak-core-based structural measurements. Section~\ref{sec:results} presents the relations between these structural diagnostics and cluster age and mass, together with the resulting two-dimensional diagnostic plane. In Section~\ref{sec:discuss}, we discuss their physical interpretation, limitations, and implications for the structural evolution of semi-resolved clusters. Section~\ref{sec:conclu} summarizes our main conclusions.

\section{Data and Methodology} \label{sec:data_method}

\subsection{Definition and Quantification of Peak Cores}
\label{subsec:datamethod_unit}

\begin{figure}[h!]
\centering
\includegraphics[width=0.5\linewidth]{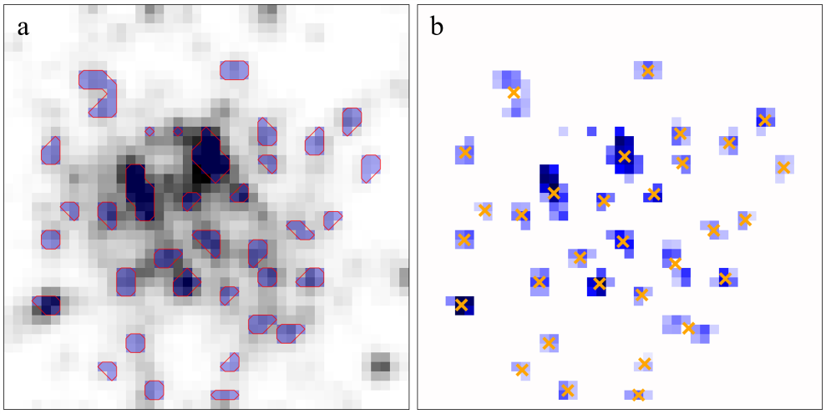}
\caption{
(a) F475W grayscale image of cluster B02-10 from our source catalog \citep{liang2026hessian}. The blue semitransparent peak cores outlined in red mark the identified peak cores.
(b) The same peak cores as in panel (a), with the blue shading indicating the value of $-\mathrm{Tr}(\mathbf{H})$ at each pixel. The pixel values within each peak core are used to define its sharpness, $S_k$. The orange crosses mark the corresponding sharpness centers, $\mathbf{r}_k$.
\label{fig:fig_1}}
\end{figure}

As discussed in Section~\ref{sec:intro} and in our previous work
\citep{liang2026hessian}, a semi-resolved cluster image can be treated as a continuous light field whose local structure is characterized by its second-order spatial derivatives, for which the Hessian matrix provides a natural representation \citep{aragon2007multiscale, schisano2014identification}:
\begin{equation}
\mathbf{H}(x,y)=
\begin{pmatrix}
\partial^{2} I/\partial x^{2} &
\partial^{2} I/\partial x\partial y \\
\partial^{2} I/\partial y\partial x &
\partial^{2} I/\partial y^{2}
\end{pmatrix},
\label{eq:hessian}
\end{equation}
where $I(x,y)$ is the surface brightness.

The signs of the two eigenvalues of $\mathbf{H}$ provide a local second-order classification of the light field. When both eigenvalues are negative, the Hessian is negative definite and the field is locally concave in every direction. In a stellar light field, such regions represent luminous entities. We therefore regard these negative-definite regions as the elementary structural units of the light field and refer to them as \emph{peak cores}:
\begin{equation}
\Omega_{\rm pc}
=
\left\{
i:\lambda_{1,i}<0,\ \lambda_{2,i}<0
\right\}
=
\bigcup_{k=1}^{N_{\rm pc}} R_k ,
\label{eq:peakcore}
\end{equation}
where $i$ denotes an image pixel and $R_k$ is the $k$th connected component of the negative-definite domain. As shown in Figure~\ref{fig:fig_1}a, the blue peak cores are outlined by the red boundaries and can remain distinct even when the underlying luminous structures are partially blended.

Within each peak core, the negative Hessian trace,
$-\mathrm{Tr}(\mathbf{H})=-(\lambda_1+\lambda_2)$, quantifies the strength of its local curvature. We define the \emph{sharpness} of the $k$th peak core as the mean negative trace within $R_k$,
\begin{equation}
S_k
=
\frac{1}{N_k}
\sum_{i\in R_k}
\left[-\mathrm{Tr}(\mathbf{H}_i)\right],
\label{eq:sharpness}
\end{equation}
where $N_k$ is the number of pixels in $R_k$. A larger $S_k$ therefore corresponds to a sharper peak core. The pixel-scale negative-trace distribution is shown as the blue spots in Figure~\ref{fig:fig_1}b.

We further define the spatial location of each peak core by its
\emph{sharpness center}, i.e., the negative-trace-weighted centroid,
\begin{equation}
\mathbf{r}_k
=
\frac{
\displaystyle\sum_{i\in R_k}
\left[-\mathrm{Tr}(\mathbf{H}_i)\right]\mathbf{r}_i
}{
\displaystyle\sum_{i\in R_k}
\left[-\mathrm{Tr}(\mathbf{H}_i)\right]
},
\label{eq:sharpness_center}
\end{equation}
where $\mathbf{r}_i=(x_i,y_i)$ is the position of pixel $i$. This quantity provides the effective spatial position of the peak core and is marked by the orange crosses in Figure~\ref{fig:fig_1}b.

The corresponding continuous formulations of $S_k$ and $\mathbf{r}_k$, together with the mathematical justification of these definitions and the derivation of the relation between $S_k$ and the intensity of an ideal Gaussian source, are presented in Appendix~\ref{sec:app_1}.

\subsection{Structural Diagnostic Indices Based on Peak Cores} \label{subsec:datamethod_index}

Based on the sharpness $S_k$ and position $\mathbf{r}_k$ defined in Section~\ref{subsec:datamethod_unit}, we construct two cluster-level diagnostics. Only peak cores whose centers lie within the cataloged effective radius of each cluster are used. First, we define the internal sharpness contrast among the peak cores as
\begin{equation}
C_{S} = \frac{\log \left(S_{\max}/S_{\min}\right)}
{1+\log \left(S_{\max}/S_{\min}\right)},
\end{equation}
where $S_{\max}$ and $S_{\min}$ are the largest and smallest core sharpness values in the cluster, respectively. As a normalized quantity, $C_{S}$ provides a convenient measure for comparing different clusters: it approaches unity when the internal sharpness contrast is large and becomes zero when all peak cores have identical sharpness. As demonstrated in Appendix~\ref{sec:app_1}, in the ideal limit, the peak core sharpness is positively correlated with the maximum of the corresponding intensity. Therefore, the level of surface-brightness fluctuation within a cluster can be characterized by $C_S$, as reflected in the variation of peak-core sharpness.

Second, we define a positional $Q$ parameter,
\begin{equation}
Q^{\rm raw}
= 
\frac{\langle m \rangle}{\langle s \rangle},
\end{equation}
where $\langle m \rangle$ is the mean edge length of the minimum spanning tree and $\langle s \rangle$ is the mean pairwise separation among all peak core centroids. Both $\langle m \rangle$ and $\langle s \rangle$ are calculated directly from the centroid coordinates $\{\mathbf{r}_k\}$.

This definition is inspired by the classical normalized $Q$ parameter \citep{cartwright2004statistical}, here denoted by $Q^{\rm norm}$, which is commonly used to characterize the spatial uniformity of a point set with well-defined source positions. To distinguish our definition from the classical one, we denote the quantity introduced here by $Q^{\rm raw}$. For a point set of fixed spatial extent, the two quantities are related as $Q^{\rm raw}\propto \rho_{\rm core}^{-1/2}Q^{\rm norm}$, where $\rho_{\rm core}$ is the surface number density of peak cores. Thus, $Q^{\rm raw}$ depends on both $Q^{\rm norm}$ and $\rho_{\rm core}$, reflecting the combined effects of spatial non-uniformity and source crowding. In the following discussion, unless otherwise specified, $Q$ denotes the raw quantity $Q^{\rm raw}$, while the normalized form is explicitly written as $Q^{\rm norm}$. A more detailed derivation is presented in Appendix~\ref{sec:app_B}.

\subsection{PHAT/AP Cluster Images and Denoising Performance} \label{subsec:datamethod_data}

The structural diagnostics introduced in Section~\ref{subsec:datamethod_index} are defined for idealized continuous fields and CCD images. In real observations, however, cluster photometric images are noisy, so additional image-processing treatment is required. To better recover the underlying stellar signal, the basic observational characteristics of the PHAT/AP cluster images must be examined.

The photometric images are drawn from the \textit{HST} Panchromatic Hubble Andromeda Treasury (PHAT; \citealt{dalcanton2012panchromatic}) survey, which imaged roughly one-third of the star-forming disk of M31 in six broadband filters from the near-ultraviolet to the near-infrared. Specifically, the data used in this work include the F275W, F336W, F475W, F814W, F110W, and F160W bands. Across these six bands, the point-spread function (PSF) core corresponds to a characteristic width of roughly $\sigma \sim 0.8$--$1$ pixel in the images used here.

The cluster sample is drawn from the Andromeda Project (AP) catalog \citep{johnson2015phat}, which provides the celestial coordinates and effective radii of the clusters. Using the catalog coordinates as cutout centers, we extract image cutouts for each cluster from the PHAT mosaics. In practice, the cutouts have a side length of about 200 pixels in the ultraviolet bands and about 160 pixels in the other bands. These sizes are sufficient to cover the full extent of a typical AP cluster and its immediate surroundings. For physical reference parameters, we adopt the CMD-based age and mass estimates from \citet{johnson2016panchromatic}.

The image-denoising and peak-core extraction procedure was applied to all 1249 AP clusters with available age and mass estimates and can be summarized as follows:

\begin{enumerate}
    \item Starting from the intensity image (Figure~\ref{fig:fig_1}a), we compute the Hessian matrix of the discrete photometric field using Gaussian-derivative filtering \citep[e.g.,][]{lindeberg1994scale} with $\sigma = 1$ pixel, which is comparable to the typical width of a point source in the \textit{HST} images. The peak-core region is then defined as the negative definite of the Hessian. Its strength is the negative sum of its two eigenvalues.

    \item To suppress noise, we compare the characteristic areas of signal and noise peak cores and reject those smaller than 4 pixels as likely noise fluctuations. The remaining peak cores are retained as the effective peak-core set used for the structural diagnostics, as illustrated in Figure~\ref{fig:fig_1}b.
\end{enumerate}

\section{Results} \label{sec:results}

\subsection{Recovery of peak-core structures and second-order structural indices with mock data} \label{subsec:result_mockdata}

In Section~\ref{subsec:datamethod_data}, we illustrated the peak-core extraction and structural-cleaning procedure using real PHAT/AP cluster images. To assess whether the extracted peak cores reliably trace genuine source structures, we construct noisy mock photometric images with known ground truth, apply the same denoising procedure, and compare the recovered peak cores with their true counterparts.

Each mock image has a size of $150\times150$ pixels and contains 5--49 compact Gaussian components, each with a fixed width of $\sigma_{\rm PSF}=1$ pixel. The component peak intensities are drawn from a lognormal distribution, with the logarithmic scatter, $\sigma_{\rm sig}$, varied from 0.1 to 2.0 in steps of 0.02. 

To generate different crowding levels, we first place the components randomly within the image and then shift them toward the image center in nine steps, each equal to one-tenth of the component's initial distance from the center. This produces 10 configurations for each realization, ranging from the initial dispersed distribution to progressively more centrally concentrated states.
We characterize the crowding of each input point set by
\begin{equation}
    C_{r,\mathrm{dot}}
    =
    \frac{\sigma_{\rm PSF}}
    {\langle d_{\rm NN}\rangle},
\end{equation}
where $\langle d_{\rm NN}\rangle$ is the mean nearest-neighbor separation of the input components. Larger values of $C_{r,\mathrm{dot}}$ therefore correspond to stronger overlap among neighboring source profiles. Combining 45 component-number values, 96 values of $\sigma_{\rm sig}$, and 10 crowding stages yields a total of 43,200 mock realizations. 

For each noise-free realization, the noise level is scaled to the faintest input component. We define the corresponding control intensity as $I_{\rm ctrl}=\min_i(A_i)$, where $A_i$ is the peak intensity of the $i$th Gaussian component. A spatially uniform sky background is then set to $I_{\rm sky}=0.7I_{\rm ctrl}$. Photon-counting fluctuations from both the sources and the sky background are introduced by pixel-wise Poisson sampling of the combined source-plus-sky intensity field. Finally, a zero-mean Gaussian read-noise component with a standard deviation of $0.35\sqrt{I_{\rm sky}}$ is added to each pixel.

\begin{figure}[h!]
\centering
\includegraphics[width=1\linewidth]{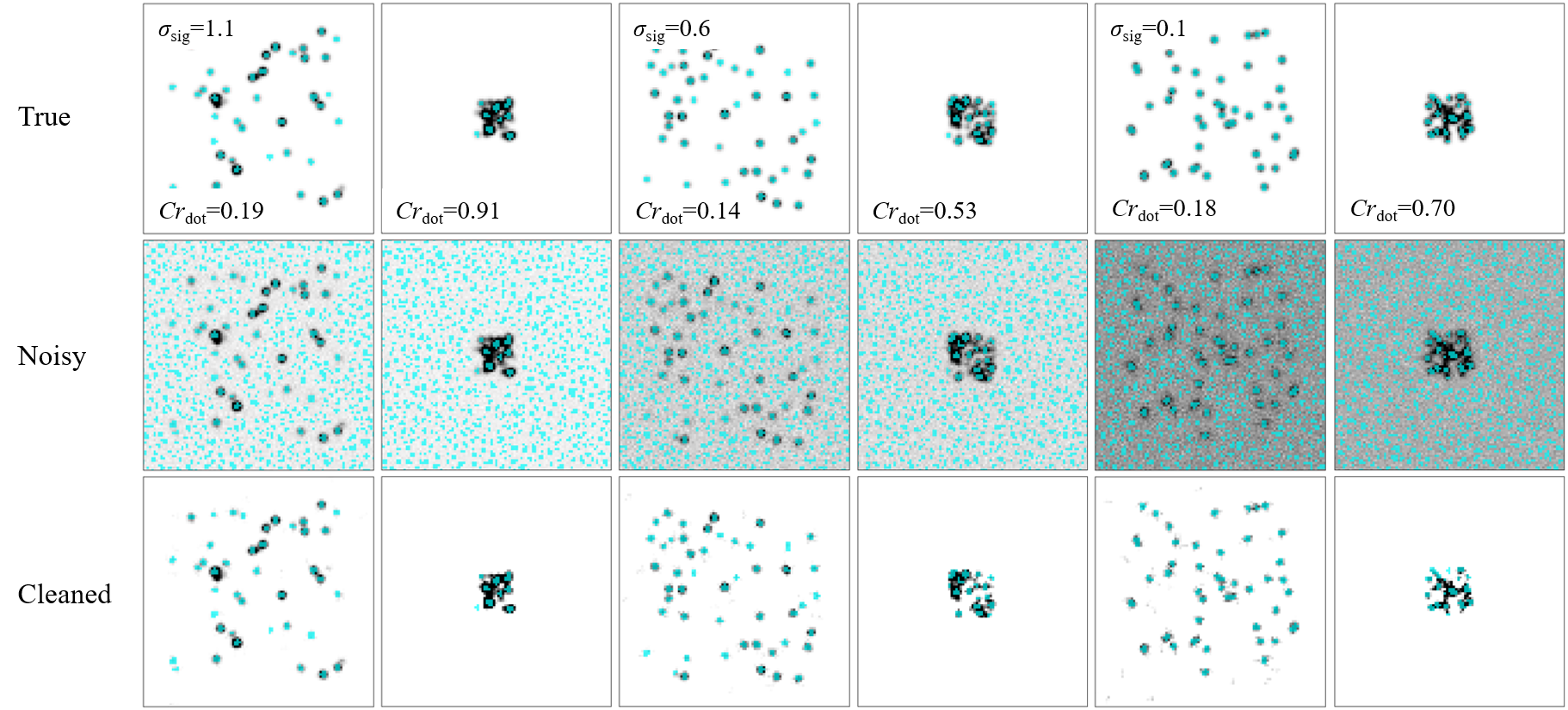}
\caption{Representative examples of peak-core recovery in the mock photometric images. The grayscale background shows the photometric intensity distribution, and the cyan regions mark the identified peak-core structures. The top, middle, and bottom rows show the true noise-free images, the corresponding noisy realizations, and the structurally cleaned images, respectively. The three pairs of columns correspond to intrinsic source-intensity scatters of $\sigma_{\rm sig}=1.1$, 0.6, and 0.1; within each pair, the left and right columns show lower- and higher-crowding configurations, respectively, with the corresponding $C_{r,\mathrm{dot}}$ values labeled in the panels.
\label{fig:fig_2}}
\end{figure}

Figure~\ref{fig:fig_2} illustrates the recovery of peak-core structures in representative mock images. The true images in the top row show that individual source components are associated with distinct peak-core structures across both contrast and crowding levels. Even in the highly crowded case with $Cr_{\mathrm{dot}}=0.91$ (second column), the peak-core boundaries remain clearly separated rather than merging into a single structure. In the noisy images (middle row), numerous small and spatially scattered peak cores appear throughout the field. This behavior is expected because the second-order derivatives are particularly sensitive to high-frequency noise, so pixel-scale fluctuations can generate abundant spurious curvature structures and prevent direct structural measurements from the noisy images. After structural cleaning (bottom row), these noise-induced features are largely removed and the underlying peak-core patterns are recovered. The recovered structures closely resemble their true counterparts over the full range of source contrasts and crowding levels illustrated here.


\begin{figure}[h!]
\centering
\includegraphics[width=0.8\linewidth]{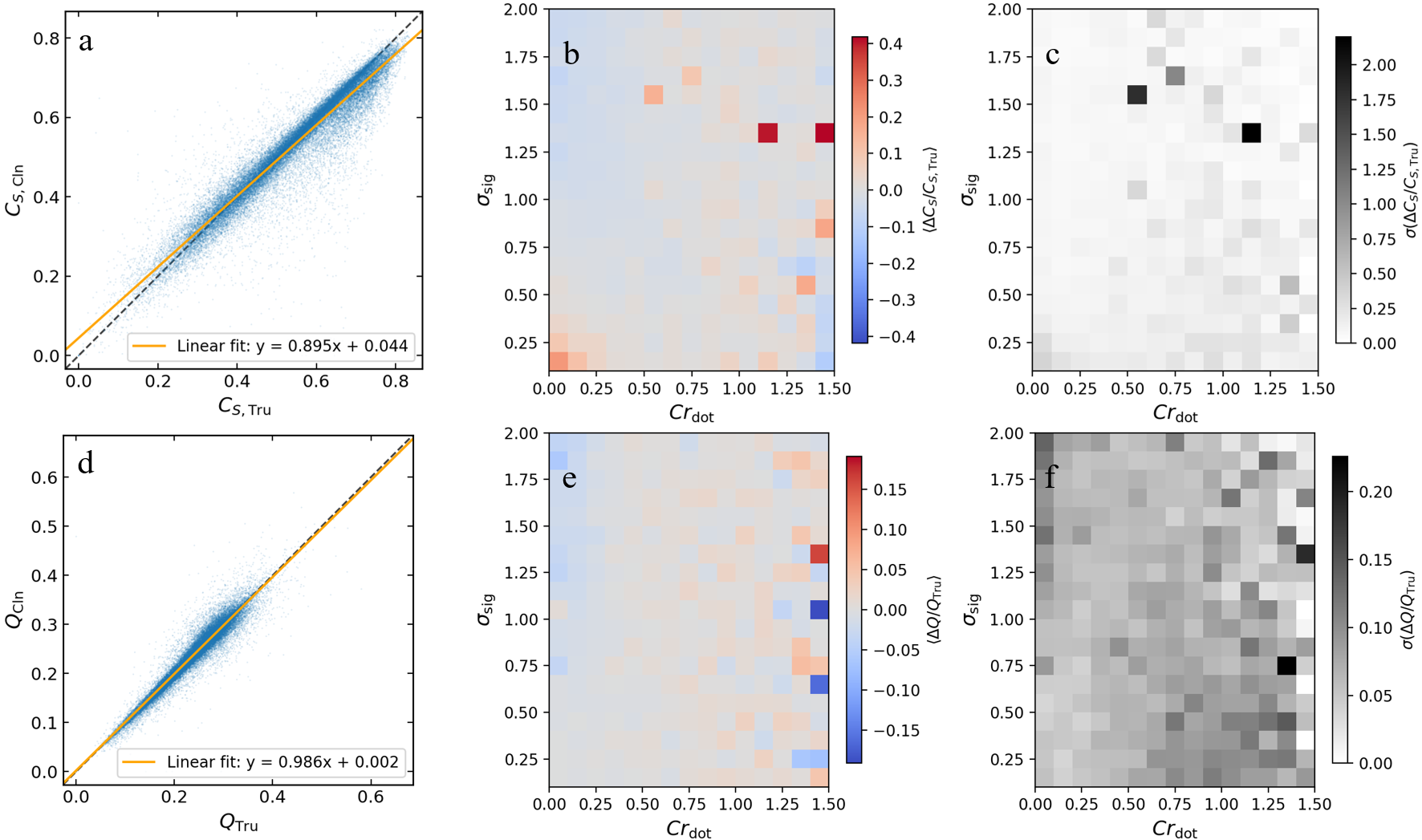}
\caption{
Recovery of $C_S$ and $Q$ from the cleaned mock maps. Panels (a)--(c) show the recovery of $C_S$, while panels (d)--(f) show that of $Q$. Panels (a) and (d) present the point-to-point comparison between the true and cleaned measurements, with each blue point representing one mock realization. The black dashed lines denote the one-to-one relations, and the orange solid lines show the linear fits, with the best-fitting relations indicated in the corresponding panels. Panels (b) and (e) show the mean normalized residual, $\langle\Delta X/X_{\rm Tru}\rangle$, in bins of $(C_{r,\rm dot},\sigma_{\rm sig})$, with a bin width of 0.1 for both parameters. Red and blue colors indicate positive and negative mean residuals, respectively. Panels (c) and (f) show the corresponding bin-wise standard deviation, $\sigma(\Delta X/X_{\rm Tru})$, where darker grayscale indicates larger scatter. Here, $X$ denotes either $C_S$ or $Q$, and $\Delta X\equiv X_{\rm Cln}-X_{\rm Tru}$.
\label{fig:fig_3}}
\end{figure}

Figure~\ref{fig:fig_3}a-c shows the recovery of $C_S$ after the cleaning procedure. As shown in panel (a), the recovered $C_S$ follows the true values closely over the full sampled range. The linear fit, $C_{S,\rm Cln}=0.895\,C_{S,\rm Tru}+0.044$, is close to the one-to-one relation, with a mild compression of the recovered dynamic range. Panel (b) shows the mean fractional residual, defined as $\Delta C_S=(C_{S,\rm Cln}-C_{S,\rm Tru})/C_{S,\rm Tru}$, across the $(C_{r,\rm dot},\sigma_{\rm sig})$ parameter space. For 90\% of the populated bins, the mean residual lies between $-5.0\%$ and $+4.7\%$, indicating only small systematic deviations over most of the parameter space. Panel (c) shows the corresponding bin-wise scatter. The median scatter is $9.2\%$, while 95\% of the populated bins have a scatter below $26.4\%$, showing that the recovery remains relatively stable across most of the explored parameter space.

Figure~\ref{fig:fig_3}d-f shows the recovery of $Q$ after the cleaning procedure. As shown in panel (d), the cleaned measurements closely follow the true values over the full range of $Q$. The linear fit, $Q_{\rm Cln}=0.986\,Q_{\rm Tru}+0.002$, is nearly consistent with the one-to-one relation, indicating an accurate recovery of the overall trend. Panel (e) shows the mean fractional residual, defined as $\Delta Q/Q_{\rm Tru}$, across the $(C_{r,\rm dot},\sigma_{\rm sig})$ parameter space. The residuals remain close to zero throughout most of the parameter space. For 90\% of the populated bins, the mean residual is within $-2.35\%$ and $+2.46\%$, indicating negligible systematic offsets in the recovered $Q$ values. Panel (f) shows the corresponding bin-wise scatter of the fractional residual. The median scatter is $6.33\%$, while 95\% of the populated bins have a scatter below $11.69\%$. 

In general, Figures~\ref{fig:fig_3} demonstrate that both $C_S$ and $Q$ are recovered with small systematic biases and limited scatter over most of the explored parameter space, indicating that the cleaning procedure effectively suppresses noise-induced structures while largely preserving the underlying structural information. Panels (b) and (e) further illustrate the applicable regime of the method. At $C_{r,\rm dot}\lesssim0.3$, noticeable residuals are present: both $Q$ and $C_S$ tend to be underestimated at larger $\sigma_{\rm sig}$, while $C_S$ is overestimated at smaller $\sigma_{\rm sig}$. In contrast, the residuals are close to zero over most of the range $0.3\lesssim C_{r,\rm dot}\lesssim1$, which, by definition of $C_{r,\rm dot}$, corresponds to the quasi-resolved regime of interest here. At $C_{r,\rm dot}\gtrsim1$, the residuals become more variable, but this regime is already severely crowded, where neighboring peak cores can become blended and structurally degenerate. Overall, the results confirm that the intrinsic structures are reliably recovered within the quasi-resolved regime relevant to this work.

\subsection{Peak-core $C_S$ and $Q$ in Young AP Clusters: Relations with Age and Mass and the $C_S$--$Q$ Diagnostic Plane} \label{subsec:result_phatdata}


\begin{figure}[h!]
\centering
\includegraphics[width=1\linewidth]{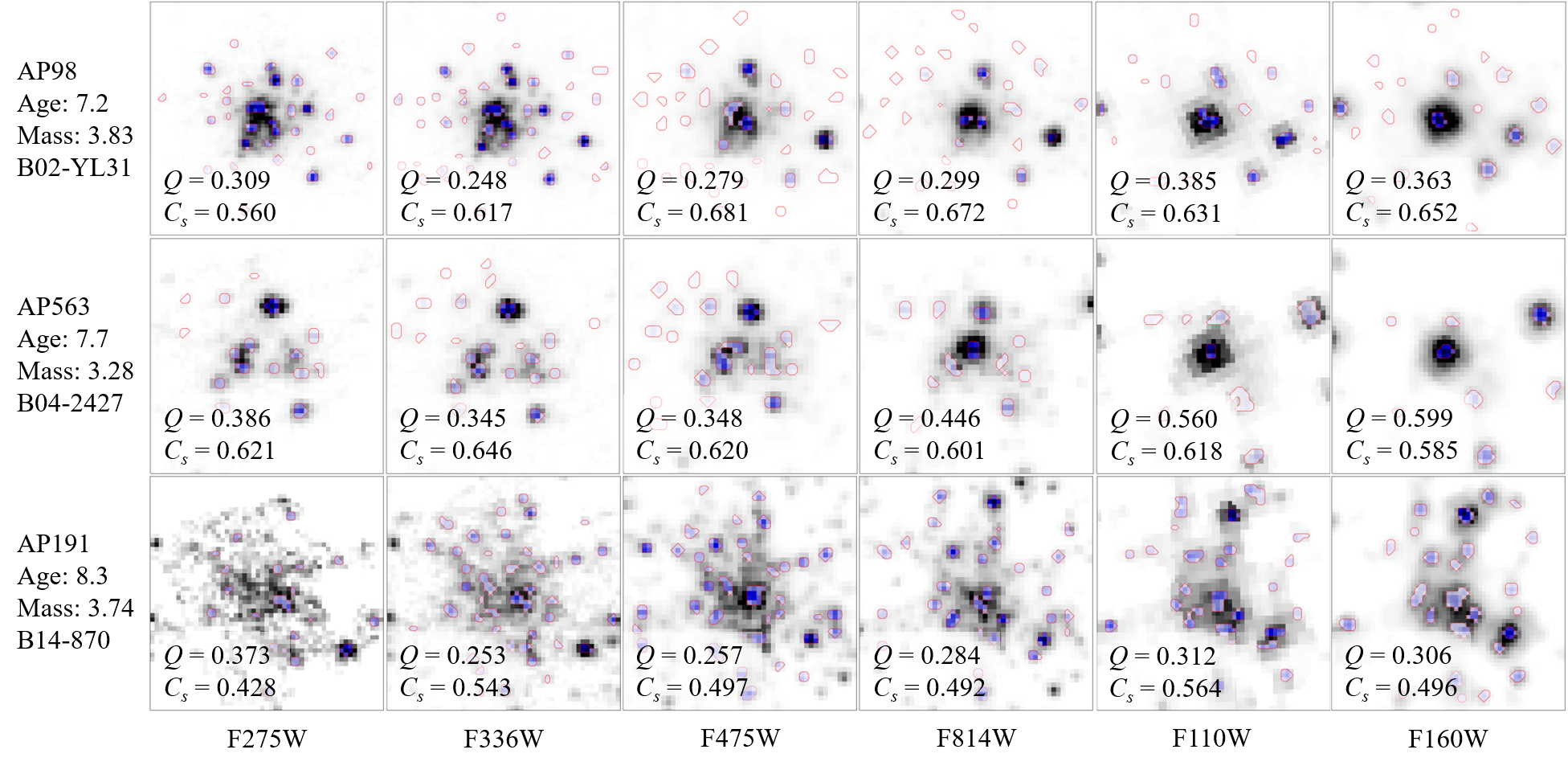}
\caption{
Representative multi-band photometric images with the identified peak cores overlaid for young AP clusters at different evolutionary stages. The grayscale backgrounds show the observed photometric intensity distributions, while the colored markers and contours indicate the locations of the identified peak cores. Each row corresponds to a cluster with a different age, and each column represents a different photometric band. The values of $C_S$ and $Q$ measured from the corresponding peak-core distributions are indicated in each panel.
\label{fig:fig_4}}
\end{figure}

The mock-data tests demonstrate the effectiveness of the denoising procedure and the ability of the recovered structural indices to trace the intrinsic properties of the underlying point set. We therefore apply the same analysis to the observed AP cluster sample. The cluster cutout images are constructed from the multiband PHAT imaging as described in Section~\ref{subsec:datamethod_data}. We exclude samples with fewer than four detected peak regions, since one to three peaks do not provide sufficient information to characterize a spatial distribution, whereas four peaks represent the minimum sampling required to define a basic structure. The resulting numbers of valid clusters in F275W, F336W, F475W, F814W, F110W, and F160W are 396, 812, 1049, 982, 947, and 898, respectively. The substantially smaller sample in F275W is mainly due to its lower signal-to-noise ratio, which reduces the number of clusters with reliably measurable structural indices.

Figure~\ref{fig:fig_4} illustrates the internal peak-core structures of three representative AP clusters spanning different age ranges. Although the clusters are only quasi-resolved and their stellar light forms a largely continuous field, the local peak cores remain spatially distinct and can be separated as individual internal structures. More importantly, the identified peak cores closely follow the observed photometric morphology: changes in the location, prominence, number, and spatial arrangement of the photometric peaks are reflected by corresponding changes in the recovered peak-core distributions. For a given cluster, these structures also vary systematically among photometric bands, indicating that the peak-core morphology is sensitive to the wavelength-dependent spatial distribution of the underlying stellar populations.


\begin{figure}[h!]
\centering
\includegraphics[width=1\linewidth]{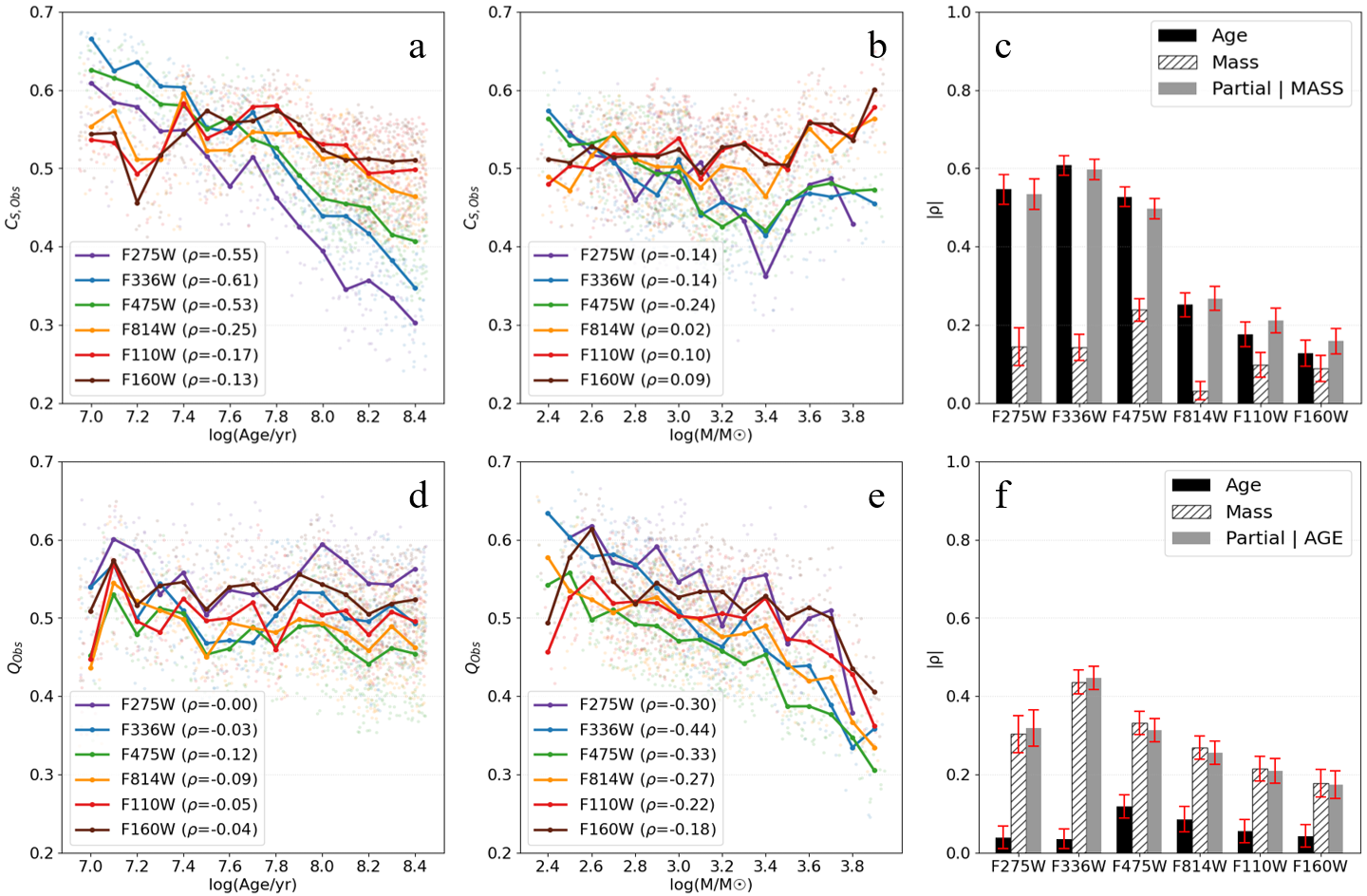}
\caption{
Spearman-rank analysis of the two structural indices against cluster age and mass for the PHAT/AP cluster sample with CMD-based ages and masses from \citet{johnson2016panchromatic}. Panels (a) and (b) show $C_{S,\rm Obs}$ as a function of $\log(\mathrm{Age/yr})$ and $\log(M/M_\odot)$, respectively, while panels (d) and (e) show the corresponding relations for $Q_{\rm Obs}$. In panels (a), (b), (d), and (e), the faint points represent individual clusters, and the solid colored curves show the median trends in running bins; the corresponding Spearman coefficients, $\rho$, are listed in the legends. Panels (c) and (f) summarize the absolute correlation coefficients, $|\rho|$, for the six photometric bands. In panel (c), the black, hatched, and gray bars denote the correlations of $C_{S,\rm Obs}$ with age, with mass, and the partial correlation with age after controlling for mass, respectively. In panel (f), the black, hatched, and gray bars denote the correlations of $Q_{\rm Obs}$ with age, with mass, and the partial correlation with mass after controlling for age, respectively. Red error bars indicate the corresponding uncertainties. The six bands, F275W, F336W, F475W, F814W, F110W, and F160W, are shown in purple, blue, green, orange, red, and brown, respectively.
\label{fig:fig_5}}
\end{figure}

The wavelength- and cluster-dependent variations in peak-core morphology illustrated above raise a natural question: whether these internal structural differences are systematically related to the physical properties of the clusters. Our previous work \citep{liang2026hessian} provided a preliminary indication that peak-core structures may contain such physical information, while the mock tests presented above establish that the corresponding structural indices can be recovered reliably over the regime relevant to the observed sample. We therefore examine systematically how the two structural indices are related to the most direct available cluster properties, namely CMD-based age and mass.

As shown in Figure~\ref{fig:fig_5}a, $C_{S,\mathrm{Obs}}$ declines systematically with increasing age in the short-wavelength bands. In F275W, F336W, and F475W, the median relations decrease from $C_{S,\mathrm{Obs}}\sim0.6$--$0.7$ at $\log(\mathrm{Age/yr})\approx7.0$ to $C_{S,\mathrm{Obs}}\sim0.3$--$0.4$ at the oldest ages, with Spearman coefficients of $\rho=-0.55$, $-0.61$, and $-0.53$, respectively. This age dependence becomes much weaker toward longer wavelengths: in F814W, F110W, and F160W, the median curves remain near $C_{S,\mathrm{Obs}}\sim0.5$--$0.6$, with weaker correlations of $\rho=-0.25$, $-0.17$, and $-0.13$. The contrast between the short- and long-wavelength bands therefore shows that the age sensitivity of $C_{S,\mathrm{Obs}}$ is itself strongly wavelength dependent. By contrast, the dependence of $C_{S,\mathrm{Obs}}$ on mass is weak overall (Figure~\ref{fig:fig_5}b), with $|\rho|\sim0.02$--$0.24$ across the six bands. Figure~\ref{fig:fig_5}c further shows that the partial correlation between $C_{S,\mathrm{Obs}}$ and age remains strong after controlling for mass, indicating that the observed age dependence is not driven solely by the covariance between age and mass.

In contrast, $Q_{\mathrm{Obs}}$ shows little dependence on age in any band. Both the median trends in Figure~\ref{fig:fig_5}d and the corresponding Spearman coefficients indicate that the relation between $Q_{\mathrm{Obs}}$ and age is weak, with $|\rho|\lesssim0.12$ throughout. However, Figure~\ref{fig:fig_5}e shows a clear negative correlation between $Q_{\mathrm{Obs}}$ and cluster mass across all bands. The trend is strongest in F336W, where $\rho=-0.44$, and remains negative in the other bands at the level of $\rho\approx-0.18$ to $-0.33$. Figure~\ref{fig:fig_5}f shows that this mass dependence persists after controlling for age, with partial correlations comparable to the original mass correlations. Thus, unlike the strongly wavelength-dependent age relation of $C_{S,\mathrm{Obs}}$, the association between $Q_{\mathrm{Obs}}$ and mass is retained across the full wavelength range considered here.

Taken together, these results show that the two second-order structural indices exhibit distinct empirical sensitivities to cluster properties: $C_{S,\mathrm{Obs}}$ is more strongly associated with age, particularly in the UV and blue bands, whereas $Q_{\mathrm{Obs}}$ shows a stronger association with mass than with age. This complementary behavior motivates a joint examination of the two indices in the $(C_{S,\rm Obs},\,Q_{\rm Obs})$ plane.


\begin{figure}[h!]
\centering
\includegraphics[width=1\linewidth]{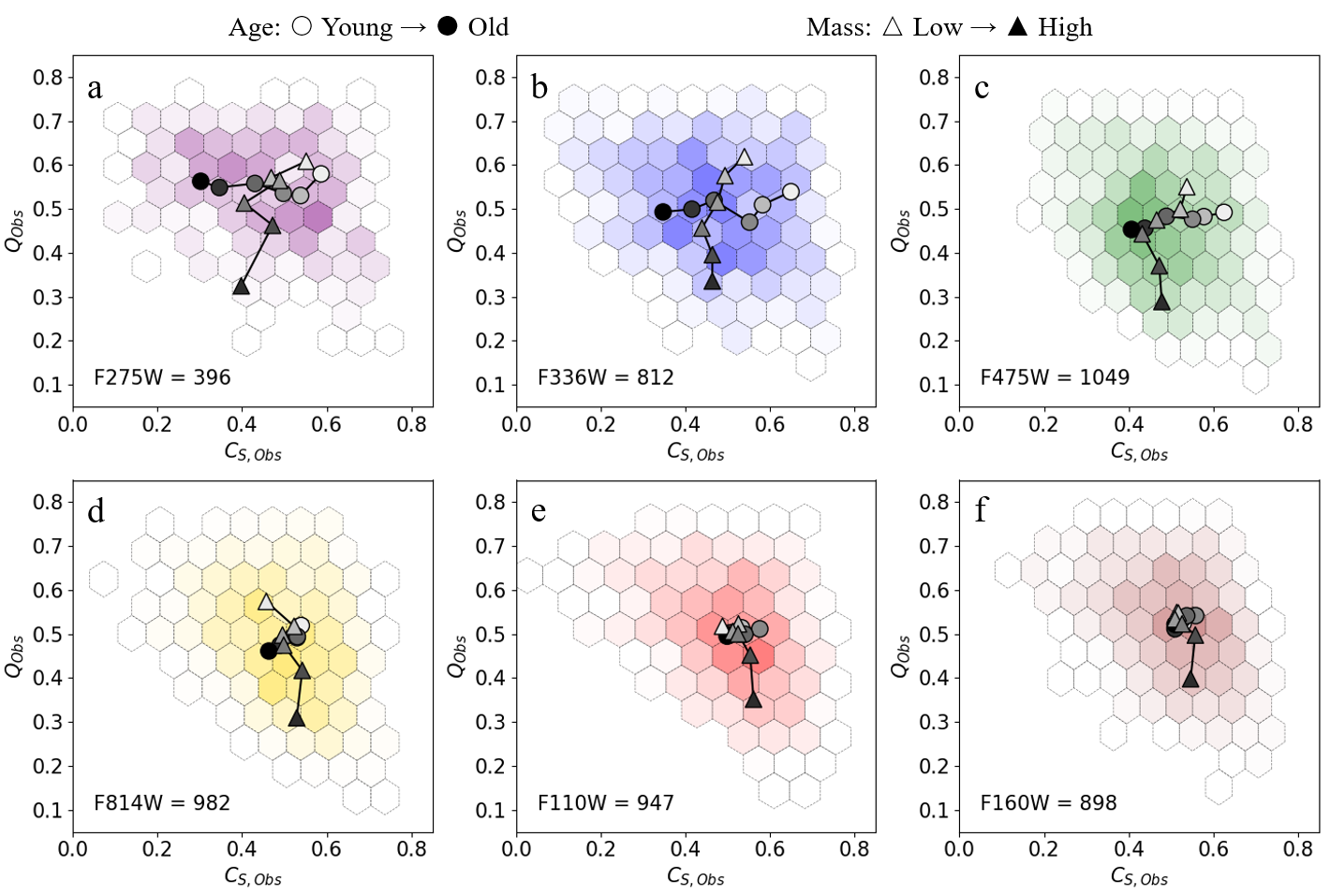}
\caption{Two-dimensional structural diagnostic plane defined by $C_{S,\rm Obs}$ and $Q_{\rm Obs}$ for the PHAT/AP cluster sample in six photometric bands: (a) F275W, (b) F336W, (c) F475W, (d) F814W, (e) F110W, and (f) F160W. The background hexagonal bins show the observed distribution of clusters in each band, with darker shading indicating a higher relative number density. Circles and triangles mark the median locations of clusters in age and mass bins, respectively. The age sequence consists of six bins over $7.0 \leq \log(\mathrm{Age/yr}) \leq 8.5$, with approximate bin boundaries at 7.0, 7.3, 7.5, 7.8, 8.0, 8.3, and 8.5, while the mass sequence consists of six bins over $2.2 \leq \log(M/M_\odot) \leq 4.5$, with approximate bin boundaries at 2.2, 2.6, 3.0, 3.4, 3.7, 4.1, and 4.5. Symbol shading progresses from light to dark with increasing age or mass. Line segments connect adjacent binned medians as guides to the eye and are not intended to represent evolutionary tracks of individual clusters.
\label{fig:fig_6}}
\end{figure}

Figure~\ref{fig:fig_6} shows the two-dimensional structural diagnostic plane for the six PHAT bands. The background hexagonal maps provide a qualitative representation of the observed cluster distribution, while the circles and triangles trace the median positions of clusters in successive age and mass bins, respectively. Across all six bands, the clusters occupy a restricted and nonuniform region of the $(C_{S,\rm Obs},\,Q_{\rm Obs})$ plane, with the highest densities generally concentrated around $C_{S,\rm Obs}\sim0.5$--$0.6$ and $Q_{\rm Obs}\sim0.4$--$0.55$. The lower-left region is sparsely populated, producing an overall wedge-like distribution. The spread is broader in the shorter-wavelength bands, particularly along the $C_{S,\rm Obs}$ direction, whereas the redder bands show a more concentrated central distribution. Thus, the observed clusters neither fill the diagnostic plane uniformly nor collapse onto a narrow one-dimensional sequence. 

More importantly, the age- and mass-binned loci reveal that this structured distribution is organized differently with respect to the two cluster properties. In the UV and blue bands, increasing age primarily shifts the median locus toward lower $C_{S,\rm Obs}$ while producing relatively little change in $Q_{\rm Obs}$, whereas increasing mass mainly drives the locus toward lower $Q_{\rm Obs}$. This separation is particularly clear in F336W and F475W, where the two loci are approximately transverse to one another. F275W exhibits the same general behavior but with a less regular mass sequence. Toward longer wavelengths, the age-binned locus becomes increasingly compact, consistent with the weaker age dependence of $C_{S,\rm Obs}$, while the mass sequence retains a more pronounced variation in $Q_{\rm Obs}$. The distinct age- and mass-binned loci highlight the complementary sensitivities of $C_{S,\rm Obs}$ and $Q_{\rm Obs}$ to cluster age and mass, respectively.

To quantify the separation of the age- and mass-dependent loci in the $(C_{S,\rm Obs},\,Q_{\rm Obs})$ plane, we define two end-to-end difference vectors, $\boldsymbol{\Delta}_{\rm age}^{\rm med}$ and $\boldsymbol{\Delta}_{\rm mass}^{\rm med}$, from the corresponding binned median sequences. Their extents define an effective rectangular area, $A_{\rm med}$, while $A_{80}$ is defined from the 5th--95th percentile ranges of $C_{S,\rm Obs}$ and $Q_{\rm Obs}$ for the full sample as a reference scale of the overall structural distribution. The coverage rate is then
\begin{equation}
R_{\rm cov}
=
\left(
\frac{A_{\rm med}}{A_{80}}
\right)^{1/2}
\end{equation}
The orthogonality rate is defined as
\begin{equation}
R_{\rm orth}
=
\left|
\sin\theta_{\rm age}\,
\sin\theta_{\rm mass}\,
\sin\theta_{\rm am}
\right|^{1/3}
\end{equation}
where $\theta_{\rm mass}$ is the angle between $\boldsymbol{\Delta}_{\rm mass}^{\rm med}$ and the horizontal direction, $\theta_{\rm age}$ is the angle between $\boldsymbol{\Delta}_{\rm age}^{\rm med}$ and the vertical direction, and $\theta_{\rm am}$ is the angle between the two vectors. A more intuitive geometric illustration, together with the explicit vector construction, is provided in Appendix~\ref{sec:app_2}.

\begin{table}[ht!]
\centering
\caption{Geometric separability metrics of the observational diagnostic plane.}
\label{tab:diag_metrics}
\begin{tabular}{lcccccc}
\hline
Metric & F275W & F336W & F475W & F814W & F110W & F160W \\
\hline
$R_{\rm cov}$  & 0.70 & 0.71 & 0.61 & 0.39 & 0.31 & 0.23 \\
$R_{\rm orth}$ & 0.91 & 0.96 & 0.96 & 0.89 & 0.96 & 0.92 \\
$S_{\rm sep}$  & 0.80 & 0.83 & 0.77 & 0.59 & 0.55 & 0.46 \\
\hline
\end{tabular}
\tablecomments{
$R_{\rm cov}$ denotes the coverage rate, $R_{\rm orth}$ the orthogonality rate, and $S_{\rm sep}=\sqrt{R_{\rm cov}R_{\rm orth}}$ the overall separability score.
}
\end{table}

Table~\ref{tab:diag_metrics} summarizes the geometric separability metrics for all six bands. The short-wavelength bands exhibit higher overall separability, with F336W reaching the largest $S_{\rm sep}$. Toward the red and near-infrared bands, the reduced separability is mainly caused by the contraction of the age-dependent locus along the $C_{S,\rm Obs}$ direction, despite the generally high orthogonality between the age and mass loci.

\section{Discussion} \label{sec:discuss}

\subsection{$C_{S,\,\rm Obs}$ as an age-sensitive structural proxy} 
\label{subsec:discus_cs_age}

Previous studies have demonstrated that pixel-scale surface-brightness fluctuations in unresolved or semi-resolved stellar systems contain information about their underlying stellar populations \citep{whitmore2011using, cook2019measuring, cook2020measuring}. In our previous work \citep{liang2026hessian}, we found empirically that a Hessian-based second-order structural statistic exhibits a correlation with cluster age. This result suggested that the local curvature of the light field may retain information about the underlying stellar population. 

In the present work, we further develop this approach by applying a more rigorous mathematical segmentation of the second-order light field, establishing an explicit correspondence between individual peak cores and compact luminous structures, and replacing the previous global empirical statistic with peak-core parameters that have clearer structural meanings. This approach also reduces the sensitivity of the measurements to pixel-scale fluctuations and noise. With the sample expanded from 254 clusters in \citep{liang2026hessian} to around $10^3$ objects in the best-sampled bands, the correlation with CMD-based age becomes stronger; in F336W, for example, $|\rho|$ increases from approximately $0.53$ to about $0.6$. We believe this relation is rooted in the physical connection between local light-field structure and stellar-population evolution.

To understand the physical basis of this connection, we consider an ideal isotropic two-dimensional Gaussian profile,
\begin{equation}
I(r)=I_0\exp\left(-\frac{r^2}{2\sigma^2}\right),
\end{equation}
where $I_0$ is the peak intensity and $\sigma$ is the characteristic width. For this profile, the peak-core region defined by the negative-definite Hessian domain corresponds to $r<\sigma$. Within this region, the sharpness measured from the Hessian trace is
\begin{equation}
S=\frac{2e^{-1/2}}{\sigma^2}I_0 .
\end{equation}
The enclosed flux within the same peak-core region is
\begin{equation}
F
=
2\pi\sigma^2\left(1-e^{-1/2}\right)I_0 .
\end{equation}
Therefore, for components with comparable characteristic widths and observed under similar effective resolution, the peak-core sharpness satisfies $S\propto F$. The detailed derivation is provided in Appendix~\ref{sec:app_1}.

For a light field composed of multiple point sources, the observed intensity distribution is given by the linear superposition of individual light profiles. Since the Hessian trace is a linear operator, the sharpness associated with each peak core reflects the local intensity contrast of the corresponding component. Therefore, $C_{S,\mathrm{Obs}}$ can be regarded as a structural representation of the local luminosity distribution, while retaining information on the spatial organization of compact light components.

The CMD-based ages are derived from the multiband luminosity distribution of member stars through stellar-evolution models. Since stellar evolution changes the luminosity hierarchy among individual stars, it naturally modifies the contrast among compact peaks in the observed light field. In this sense, the age dependence of $C_{S,\mathrm{Obs}}$ originates from the same underlying stellar-population evolution that drives the multiband photometric variations used in CMD fitting. Photometry integrated within the characteristic scale $\sigma$ of individual peak cores may therefore also carry useful age information and could potentially be incorporated into CMD-based analyses. Since the present work focuses on structural diagnostics, such a photometric comparison is left for future work as an independent test of the connection between peak-core structure and stellar-population evolution.

Over the $10$--$300$ Myr age range considered here, the stronger age dependence of $C_{S,\mathrm{Obs}}$ in the UV and blue bands is naturally associated with the rapid fading and disappearance of massive stars. At longer wavelengths, contributions from stars spanning different masses and evolutionary phases, including luminous evolved stars, make the luminosity evolution less monotonic and therefore weaken the corresponding age dependence. This wavelength-dependent behavior is consistent with the trend found for the Hessian-based statistic in \citet{liang2026hessian} and suggests that the fading of massive stars is accompanied by a progressive loss of high-curvature structures in the cluster light distribution.

\subsection{$Q_{\mathrm{Obs}}$ as a practical mass-sensitive diagnostic of projected structure} \label{subsec:discus_q_mass}

The normalized $Q$ was introduced as a measure of cluster spatial structure \citep{cartwright2004statistical,schmeja2006evolving}. Values of $Q^{\rm norm}<0.8$ are generally associated with substructured and spatially inhomogeneous configurations \citep{cartwright2004statistical,parker2012characterizing}. \citet{lomax2011statistical} further showed that $Q^{\rm norm}$ can also characterize the structure of continuous fields. Our mock tests in Section~\ref{subsec:result_mockdata} likewise show that $Q^{\rm raw}_{\mathrm{Dns}}$ is positively correlated with the precise point-source coordinates. 

We treat $Q^{\rm raw}_{\mathrm{Obs}}$ as an empirical measure of projected structure in semi-resolved cluster images, where ``structure'' refers to both the spatial arrangement and the surface number density of the peak cores. For a semi-resolved image of a cluster with $N$ detected peak cores
\begin{equation} \label{equ:qraw_qnorm}
Q^{\rm raw}
=
\frac{\sqrt{N A_{\rm eff}}}{(N-1)\,R_{\rm eff}}\,
Q^{\rm norm}
\;\approx\;
\frac{1}{\sqrt{\rho_{\rm core}}\,R_{\rm eff}}\,Q^{\rm norm},
\end{equation}
where $R_{\rm eff}$ is the half-light radius, $A_{\rm eff}$ is the area enclosed within $R_{\rm eff}$, and $\rho_{\rm core}$ is the surface number density of peak cores. In our sample, $R_{\rm eff}$ typically ranges from 1 to 2 pc, so $Q_{\mathrm{Obs}}^{\mathrm{raw}}$ is sensitive to both $Q_{\mathrm{Obs}}^{\mathrm{norm}}$ and $\rho_{\rm core}$. Moreover, $Q_{\mathrm{Obs}}^{\mathrm{norm}}$ typically ranges from 0 to slightly above 1, whereas, for the typical values of $N$ ($\sim4$--50) and $R_{\rm eff}$, $\sqrt{\rho_{\rm core}}$ varies from $\sim0.6$ to $4~\mathrm{pc}^{-1}$, spanning a broader range than $Q_{\mathrm{Obs}}^{\mathrm{norm}}$. This implies that $Q_{\mathrm{Obs}}^{\mathrm{raw}}$ may be more strongly modulated by $\rho_{\rm core}$, or more directly by $N$. A detailed derivation is provided in Appendix~\ref{sec:app_B}.

Nevertheless, $N$ is not an ideal primary structural parameter because it contains no information about the spatial configuration of the identified structures. Moreover, its correlation with cluster mass is comparable to that of $Q_{\mathrm{Obs}}^{\mathrm{raw}}$; in F336W, for example, $\rho\simeq0.47$ for $N$, compared with $|\rho|\simeq0.44$ for $Q_{\mathrm{Obs}}^{\mathrm{raw}}$. We also do not adopt $Q_{\mathrm{Obs}}^{\mathrm{norm}}$ as the primary diagnostic because it performs less well for semi-resolved images. Its strongest correlation with cluster mass is only $\sim0.41$ in F336W, while the coverage rate on the $(C_{S,\rm Obs},\,Q_{\mathrm{Obs}}^{\mathrm{norm}})$ plane is only $\sim0.61$, both lower than the corresponding values for $Q_{\mathrm{Obs}}^{\mathrm{raw}}$.

Overall, the goal of this framework is not to recover a uniquely defined structural parameter in the idealized point-source sense, but to provide a fast and practical diagnostic for semi-resolved clusters without requiring star-by-star source identification. In this context, $Q_{\mathrm{Obs}}^{\mathrm{raw}}$ remains useful because it is dimensionless, easy to measure, and jointly responsive to both components of the projected structural complexity and the surface density of detectable peak cores. 

\subsection{Physical interpretation and limitations of the diagnostic plane}
\label{subsec:discus_plane}

As shown in Section~\ref{subsec:result_phatdata}, $C_{S,\mathrm{Obs}}$ and $Q_{\mathrm{Obs}}$ correlate with cluster age and mass, respectively, and the same trends are evident in the $(C_{S,\mathrm{Obs}},\,Q_{\mathrm{Obs}})$ plane across multiple bands. The median age sequences extend primarily along the $C_{S,\mathrm{Obs}}$ direction in the blue bands, while the median mass sequences remain predominantly vertical over a broader wavelength range. These consistent behaviors across bands support the physical significance of the diagnostic plane.

An ideal diagnostic plane would separate the age and mass dependences along two nearly orthogonal directions. F336W most closely approaches this configuration (Figure~\ref{fig:fig_6}b), with the median age and mass sequences extending approximately horizontally and vertically, respectively. Although the plane does not yet provide unique age and mass estimates for individual clusters, it can provide relative information on age and mass variations for cluster populations within comparable distance and physical environments.

The age sequences in the other two short-wavelength bands also remain relatively well extended. Toward redder bands, however, the age sequences become less distinct, where luminous evolved stars, such as red supergiants and AGB stars, can emerge at different ages as the stellar population passes through short-lived evolutionary phases. The mass sequences exhibit opposite tilts across wavelength: with increasing cluster mass, $C_{S,\mathrm{Obs}}$ decreases in the blue bands but increases in the red bands. This may reflect different mass dependences of peak-core number and contrast, as illustrated in the first row of Figure~\ref{fig:fig_4}. In the blue bands, more massive clusters contain more comparable peak cores, whereas in the red bands they are more likely to host luminous evolved stars that produce prominent local peaks.

Another intriguing feature of the diagnostic plane is the sparse occupation of the lower-left region in all six bands. This region likely corresponds to relatively old and massive clusters, which are less represented in our UV-selected sample that primarily traces young OB-star clusters. In addition, the apparent gap may partly reflect observational incompleteness. Some young clusters with numerous but low-contrast peak cores may fall below the effective detection limit of HST. A second observational limitation arises from the semi-resolved nature of our method. The projected angular separations among luminous cluster members must fall within an appropriate range relative to the instrumental resolution set by the PSF. Consequently, comparisons across filters, instruments, or galaxies require further calibration for differences in effective spatial resolution and PSF characteristics. At present, the absolute values of $C_{S,\mathrm{Obs}}$ and $Q_{\mathrm{Obs}}$ should therefore not be directly compared across bands.

\section{Conclusion} \label{sec:conclu}

In this work, we further demonstrate that the physical properties of semi-resolved star clusters can be reflected by the second-order structures of their photometric images. Using the Hessian matrix, we define peak cores as the basic structural units of the cluster light distribution and apply this framework to the multiband photometric images of star clusters in M31.

Applying this framework to 1249 star clusters, we find that cluster age is primarily reflected in the sharpness and contrast of peak cores, whereas cluster mass is more closely related to their spatial organization. Combining these two structural properties further yields a diagnostic plane in which age and mass show distinct extensions, most clearly in the F336W band. This plane can therefore be used to assess the relative ages and masses of cluster samples observed under comparable environmental and observational conditions.

To validate the structural denoising method, we conducted extensive mock-image tests. The results show that the recovered structural indices closely reproduce those of the corresponding noise-free images, demonstrating the reliability of our method over the usable parameter range.

Future work will improve this framework in three directions. First, more realistic astronomical and physical constraints will be incorporated into simulations. Second, the Hessian-based framework will be extended to define a broader set of basic structural units and their relations, to build a more complete and self-consistent theoretical description of cluster structure. Third, the method will be applied to data from different telescopes and nearby galaxies to enlarge the sample coverage. In particular, the upcoming CSST mission \citep{zhan2021wide}, with wide-field imaging from the near-ultraviolet to the near-infrared, will provide valuable opportunities to extend this analysis to larger and more diverse samples of semi-resolved star clusters.

\begin{acknowledgments}

This work was supported by the National Natural Science Foundation of China (NSFC) under Grant Nos. 12588202, 12041302, and 12073038, the China Manned Space Engineering Program (China Space Station Telescope, CSST) under Grant Nos. CSST-2021-A08, CSST-2021-B02, CSST-2021-B03, and CSST-2021-B06, and the National Key R\&D Program of China under Grant No. 2023YFA1608004.

\end{acknowledgments}

\begin{contribution}

Y.Liang conceived the project, developed the methodology, carried out the analysis, interpreted the results, and wrote the manuscript. Jingwen Wu and Chao-wei Tsai provided funding support, supervised the broader project context, and commented on the manuscript.


\end{contribution}

%



\appendix \label{sec:app}

\section{Sharpness--Flux Relation in Gaussian peak cores} \label{sec:app_1}



For an ideal isolated point source, we assume a 2D isotropic Gaussian profile
\begin{equation}
I(\mathbf{x}) = I_0\exp\left[-\frac{|\mathbf{x}-\mathbf{x}_0|^2}{2\sigma^2}\right],
\end{equation}
where $\mathbf{x}_0$ is the source center, $I_0$ is the central intensity, and $\sigma$ is the Gaussian width. For convenience, we define
\begin{equation}
r \equiv |\mathbf{x}-\mathbf{x}_0|.
\end{equation}
The trace of the Hessian is then
\begin{equation}
T(\mathbf{x}) \equiv \mathrm{Tr}(H_I) = \Delta I
=\left(\frac{r^2}{\sigma^4}-\frac{2}{\sigma^2}\right)I(\mathbf{x}).
\end{equation}
For an isotropic Gaussian, the radial and tangential eigenvalues of the Hessian are
\begin{equation}
\lambda_r
=
\left(\frac{r^2}{\sigma^4}-\frac{1}{\sigma^2}\right)I(\mathbf{x}),
\qquad
\lambda_t
=
-\frac{1}{\sigma^2}I(\mathbf{x}).
\end{equation}
The Hessian is therefore negative definite for $r<\sigma$, and the peak-core boundary is given by
\begin{equation}
r=\sigma.
\end{equation}
We denote the corresponding isolated peak core by $R$, with area
\begin{equation}
A = \pi\sigma^2.
\end{equation}

If the sharpness is defined as the area-averaged negative trace within the peak core, then
\begin{equation}
S \equiv -\frac{1}{A}\int_R T(\mathbf{x})\,dA.
\end{equation}
Using $T=\Delta I$ and the divergence theorem, one obtains
\begin{equation}
\int_R T(\mathbf{x})\,dA
=
\oint_{\partial R} \frac{\partial I}{\partial n}\,ds
=
-\frac{2\pi I_0}{\sqrt{e}},
\end{equation}
and therefore
\begin{equation}
S=\frac{2}{\sqrt{e}\,\sigma^2}I_0.
\end{equation}

The flux enclosed by the isolated peak core is
\begin{equation}
F
=
\int_R I(\mathbf{x})\,dA
=
2\pi\sigma^2\left(1-e^{-1/2}\right)I_0.
\end{equation}
Eliminating $I_0$ gives
\begin{equation}
S=\frac{1}{\pi(\sqrt{e}-1)\sigma^4}F.
\end{equation}

We now extend this result to a weakly blended Gaussian point field. Let the total intensity field be
\begin{equation}
I(\mathbf{x})=\sum_k I_k(\mathbf{x}),
\end{equation}
so that, by linearity of the trace operator,
\begin{equation}
T(\mathbf{x}) \equiv \mathrm{Tr}(H_{I_{\mathrm{mix}}})=\Delta I
=\sum_k \Delta I_k.
\end{equation}
For a local peak core $R_k$ dominated by source $k$, we define the local mixed field as
\begin{equation}
I_{k,\mathrm{mix}}(\mathbf{x}) \equiv I_k(\mathbf{x})+\delta I_k(\mathbf{x}),
\qquad
\delta I_k(\mathbf{x}) \equiv \sum_{j\ne k} I_j(\mathbf{x}),
\end{equation}
and correspondingly
\begin{equation}
T_{k,\mathrm{mix}}(\mathbf{x}) \equiv T_k(\mathbf{x})+\delta T_k(\mathbf{x}),
\qquad
\delta T_k(\mathbf{x}) \equiv \sum_{j\ne k} T_j(\mathbf{x}).
\end{equation}

Under the mildly blended conditions considered here, the neighboring contribution within the local negative-definite region can be treated as a perturbation, such that
\begin{equation}
|\delta I_k(\mathbf{x})|\ll |I_k(\mathbf{x})|,
\qquad
|\delta T_k(\mathbf{x})|\ll |T_k(\mathbf{x})|
\qquad
\text{for } \mathbf{x}\in R_k.
\end{equation}
It then follows that the mixed-field sharpness and enclosed flux for the $k$th local peak core can be written as
\begin{equation}
S_{k,\mathrm{mix}} = S_k + \delta S_k,
\qquad
F_{k,\mathrm{mix}} = F_k + \delta F_k,
\end{equation}
where $\delta S_k$ and $\delta F_k$ are small perturbative corrections induced by neighboring sources. Summing over all local peak cores gives
\begin{equation}
S = \sum_k S_{k,\mathrm{mix}}
=
\sum_k S_k + \sum_k \delta S_k,
\qquad
F = \sum_k F_{k,\mathrm{mix}}
=
\sum_k F_k + \sum_k \delta F_k.
\end{equation}
Using the isolated-source relation
\begin{equation}
S_k
=
\frac{1}{\pi(\sqrt{e}-1)\sigma^4}F_k,
\end{equation}
we obtain
\begin{equation}
S = \sum_k S_{k,\mathrm{mix}}
=
\frac{1}{\pi(\sqrt{e}-1)\sigma^4}\sum_k F_k + \sum_k \delta S_k.
\end{equation}
Therefore, if the summed perturbation remains higher order under weak blending, the mixed field is approximately equivalent to the superposition of isolated-source contributions:
\begin{equation}
S
\approx
\frac{1}{\pi(\sqrt{e}-1)\sigma^4}F
\propto F.
\end{equation}
In this sense, source blending introduces only a small higher-order correction and does not destroy the approximate linear superposability of the single-source sharpness--flux relation.

In astrophysical terms, this approximation is justified because, in most semi-resolved stellar clusters considered here, the central flux of a point source is still much larger than the summed overlap flux contributed by neighboring wings. Thus, even if source blending modifies the outer peak-core morphology, it does not usually compete with the peak intensity itself. The local mixed field therefore remains perturbatively close to the isolated-source case, and the approximate proportionality between sharpness and enclosed flux is expected to hold for most sources in practice.

\section{Relation between the Raw and Normalized \texorpdfstring{$Q$}{Q} Parameters} \label{sec:app_B}

As derived in Appendix~\ref{sec:app_1}, each peak core $R_k$ is defined by a closed zero-trace contour in the trace map, and its effective position is given by the trace centroid
\begin{equation}
\mathbf{r}_k
=
\frac{\int_{R_k} \mathbf{x}\,[-T(\mathbf{x})]\, dA}
{\int_{R_k} [-T(\mathbf{x})]\, dA},
\qquad k=1,\dots,N ,
\end{equation}
where $T(\mathbf{x}) \equiv \mathrm{Tr}(H_I)$, and $N$ is the total number of peak cores. In the present work, the $Q$ parameter is constructed from the point set $\{\mathbf{r}_k\}$ rather than from individual stellar coordinates.

For any pair of peak cores, we define the separation
\begin{equation}
d_{ij} \equiv |\mathbf{r}_i-\mathbf{r}_j|,
\qquad 1\le i<j\le N .
\end{equation}
The mean pairwise separation of the complete graph is then
\begin{equation}
\langle s \rangle
=
\frac{2}{N(N-1)}\sum_{i<j} d_{ij}.
\end{equation}

Next, we construct the minimum spanning tree (MST) on the same point set $\{\mathbf{r}_k\}$. If the MST contains $N-1$ edges with lengths $\ell_1,\dots,\ell_{N-1}$, then the mean MST edge length is
\begin{equation}
\langle m \rangle
=
\frac{1}{N-1}\sum_{n=1}^{N-1}\ell_n.
\end{equation}
We define the raw $Q$ parameter as
\begin{equation}
Q^{\mathrm{raw}}
\equiv
\frac{\langle m \rangle}{\langle s \rangle}.
\end{equation}

Following Cartwright \& Whitworth (2004), the normalized mean separation is defined as
\begin{equation}
\bar{s}
=
\frac{1}{R_{\mathrm{eff}}}\langle s \rangle,
\end{equation}
where the effective radius of the point set is
\begin{equation}
R_{\mathrm{eff}}
=
\max_k |\mathbf{r}_k-\bar{\mathbf{r}}|,
\qquad
\bar{\mathbf{r}}
=
\frac{1}{N}\sum_{k=1}^{N}\mathbf{r}_k .
\end{equation}
The normalized mean MST edge length is
\begin{equation}
\bar{m}
=
\frac{N-1}{\sqrt{N A_{\mathrm{eff}}}} \langle m \rangle,
\end{equation}
where $A_{\mathrm{eff}}$ is the projected area associated with the point set. The normalized $Q$ parameter is therefore
\begin{equation}
Q^{\mathrm{norm}}
\equiv
\frac{\bar{m}}{\bar{s}}.
\end{equation}

Substituting the above definitions gives
\begin{equation}
Q^{\mathrm{norm}}
=
\frac{\langle m \rangle}{\langle s \rangle}
\cdot
\frac{(N-1)R_{\mathrm{eff}}}{\sqrt{N A_{\mathrm{eff}}}}
=
\frac{(N-1)R_{\mathrm{eff}}}{\sqrt{N A_{\mathrm{eff}}}}
\,Q^{\mathrm{raw}}.
\end{equation}

Defining the number surface density of the peak-core centroids as
\begin{equation}
\rho_{\mathrm{core}} \equiv \frac{N}{A_{\mathrm{eff}}},
\end{equation}
we obtain
\begin{equation}
Q^{\mathrm{raw}}
=
\frac{N}{(N-1)R_{\mathrm{eff}}\sqrt{\rho_{\mathrm{core}}}}
\,Q^{\mathrm{norm}}.
\end{equation}

If $N \gg 1$ and $R_{\mathrm{eff}}$ varies only weakly among the samples considered (e.g., remains within $\sim$1--2 pc), then
\begin{equation}
Q^{\mathrm{raw}} 
\approx
\frac{1}{\sqrt{\rho_{\mathrm{core}}}}\,Q^{\mathrm{norm}}.
\end{equation}
This shows that the raw $Q$ parameter is jointly determined by the normalized structural parameter and the surface number density of the peak cores. In this sense, $Q^{\mathrm{raw}}$ reflects both the uniformity and the crowdedness of the peak-core distribution.

\section{Quantifying the Separability of the $(C_S,Q)$ Plane} \label{sec:app_2}
\begin{figure}[h!]
\centering
\includegraphics[width=0.4\linewidth]{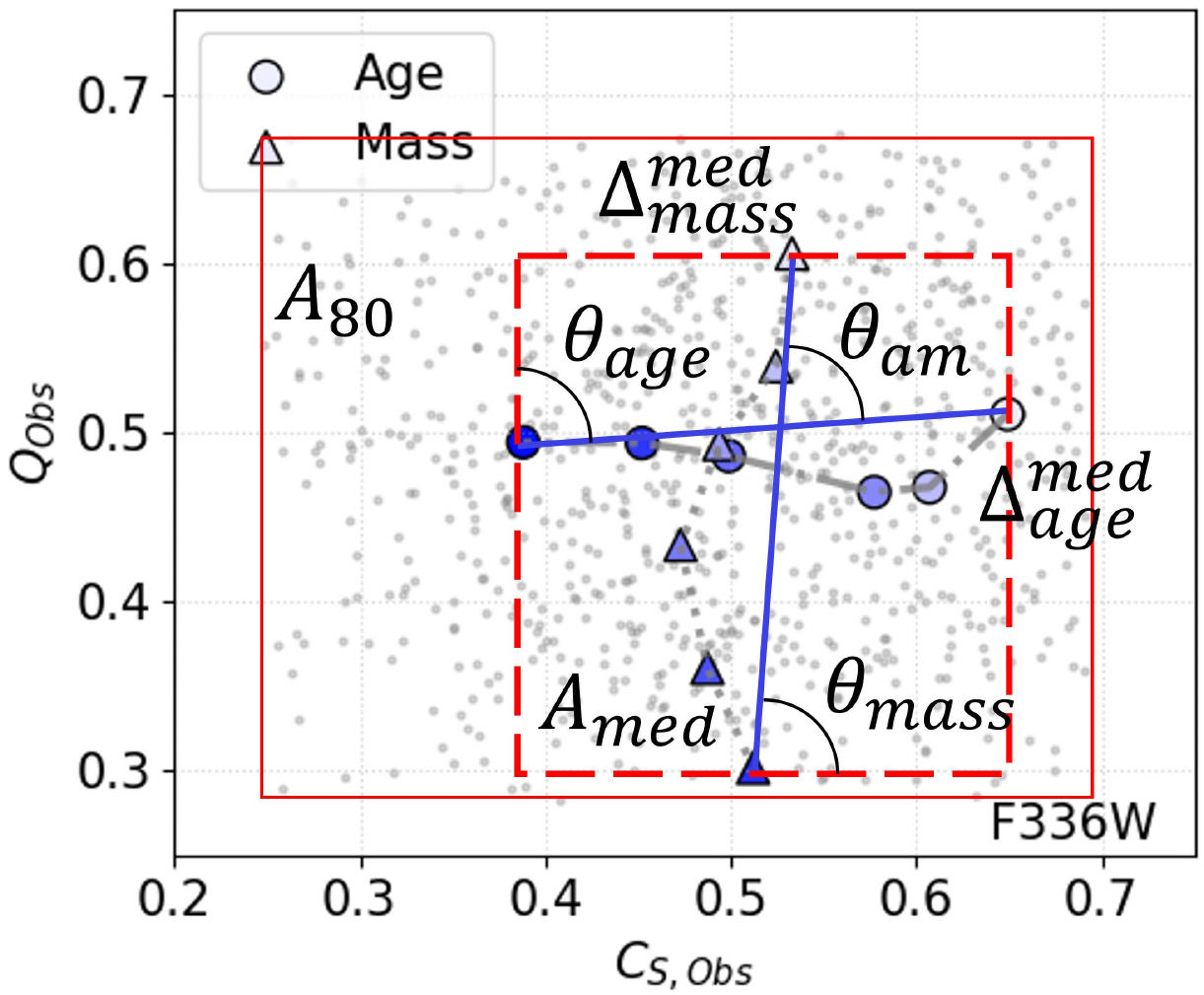}
\caption{
Illustration of the geometrical quantities used to quantify the separability of the $(C_S,Q)$ diagnostic plane, taking the F336W band as an example. The gray points denote the sample enclosed by the solid red rectangle with area $A_{80}$. The two blue line segments, $\boldsymbol{\Delta}_{\rm age}^{\rm med}$ and $\boldsymbol{\Delta}_{\rm mass}^{\rm med}$, connect the two endpoint medians of the age and mass sequences, respectively. The dashed red rectangle marks the median-spanned area $A_{\rm med}$. The three angles are defined as $\theta_{\rm mass}$, between $\boldsymbol{\Delta}_{\rm mass}^{\rm med}$ and the horizontal direction; $\theta_{\rm age}$, between $\boldsymbol{\Delta}_{\rm age}^{\rm med}$ and the vertical direction; and $\theta_{\rm am}$, between the two median segments.
\label{fig:fig_7}}
\end{figure}

To quantify the separability of the $(C_S,Q)$ diagnostic plane, we define two geometric indicators: a coverage rate and an orthogonality rate (Figure~\ref{fig:fig_7}).

For the coverage rate, we first define two end-to-end difference vectors, $\boldsymbol{\Delta}_{\rm age}^{\rm med}$ and $\boldsymbol{\Delta}_{\rm mass}^{\rm med}$. As shown in Figure~\ref{fig:fig_7}, the two blue solid lines connect the two extreme points of the age- and mass-binned median loci, respectively. We then use the horizontal span of $\boldsymbol{\Delta}_{\rm age}^{\rm med}$ and the vertical span of $\boldsymbol{\Delta}_{\rm mass}^{\rm med}$ to define an axis-aligned effective rectangular area,
\begin{equation}
A_{\rm med}=
\left|\Delta_{{\rm age},x}^{\rm med}\right|
\left|\Delta_{{\rm mass},y}^{\rm med}\right|,
\end{equation}
which characterizes the extent jointly spanned by the median age and mass trends in the diagnostic plane.

To provide a reference scale for the overall distribution, we define
\begin{equation}
R_{\rm cov}=\left(\frac{A_{\rm med}}{A_{80}}\right)^{1/2},
\end{equation}
where $A_{80}$ is an effective rectangular area constructed from the 5th--95th percentile ranges of $C_S$ and $Q$ for the full cluster sample. In other words, its horizontal and vertical sides correspond to the central 90\% ranges of the two coordinates, respectively, so that $A_{80}$ represents an effective area of about $0.9\times0.9\approx 0.81$ of the full coordinate extent, rather than the exact area enclosing 80\% of the sample points.

The orthogonality rate measures how close the two median trends are to an ideal age--mass decoupled configuration:
\begin{equation}
R_{\rm orth}=
\left|
\sin\theta_{\rm age}\,
\sin\theta_{\rm mass}\,
\sin\theta_{\rm am}
\right|^{1/3},
\end{equation}
where $\theta_{\rm mass}$ is the angle between $\boldsymbol{\Delta}_{\rm mass}^{\rm med}$ and the horizontal direction, $\theta_{\rm age}$ is the angle between $\boldsymbol{\Delta}_{\rm age}^{\rm med}$ and the vertical direction, and $\theta_{\rm am}$ is the angle between the two median segments.

Larger values of $R_{\rm cov}$ and $R_{\rm orth}$ indicate stronger separability of the diagnostic plane.


\bibliography{Reference}{}
\bibliographystyle{aasjournalv7}



\end{document}